\documentclass{iaus}
\usepackage{graphicx}

\title[SMART -- a computer program for modelling stellar atmospheres] 
{SMART -- a computer program for modelling stellar atmospheres}

\author[A. Aret, A. Sapar, R. Poolam\"ae \& L. Sapar]   
{A. Aret, A. Sapar, R. Poolam\"ae \and L. Sapar}

\affiliation{Tartu Observatory, 61602 T\~oravere, Tartumaa, Estonia \break email: aret@aai.ee}

\pubyear{2008}
\volume{252}  
\pagerange{41--42}
\date{?? and in revised form ??}
\jname{The Art of Modelling Stars in the 21$^{st}$ Century}
\editors{L. Deng \& K.L. Chan, eds.}
\doi{10.1017/S1743921308022394}
\begin{document}

\maketitle

\begin{abstract}
Program SMART (Spectra and Model Atmospheres by Radiative Transfer)
has been composed for modelling atmospheres and spectra of hot
stars (O, B and A spectral classes) and studying different
physical processes in them \cite{tubingen,sapar_moskva1}.
Line-blanketed models are computed assuming plane-parallel, static and horizontally homogeneous
atmosphere in radiative, hydrostatic and local thermodynamic equilibrium.
Main advantages of SMART are its shortness, simplicity, user friendliness and
flexibility for study of different physical processes.
SMART successfully runs on PC both under Windows and Linux.
\end{abstract}

 \firstsection 
 \section{Main features of the program}

Model atmospheres are calculated iteratively varying only
temperature and pressure dependency on column density. Flux
constancy about 0.1--0.5~\% is achieved by about 10 iterations,
using Kurucz ATLAS9 models as input. Number of atmospheric layers
can be multiplied up if necessary. Line absorption has been
completely taken into account with spectral resolution 300~000.

Programming language is Fortran 90, the program is compiled using Intel Fortran Compiler and runs both on Windows
and Linux computers. Code is extensively commented on right-hand margin.
Graphical interface (written in C++) has been composed for visualizing results of calculation.

Radiative transfer has been calculated using integration by parts, yielding series of exponential integrals.
The scattering processes are computed by simple $\Lambda$-iteration.
Radiative transfer calculations give radiative flux $F_{\nu}(\lambda,\tau)$ in all layers of atmosphere.

Capabilities of program SMART include also computations of evolution of diffusive separation of isotopes in
atmospheres of CP stars, relaxational formation of NLTE in line spectra, accelerations
of clumps in stellar wind, computation of detailed spectral limb darkening and
hence the spectra of rotating stars and  non-irradiated eclipsing binaries.
Pan-spectral method for determining element abundances from high-quality observed spectra have been developed and
implemented as an extension to SMART.

The basic restriction is the assumption of plane-parallel, static and horizontally homogeneous
atmosphere with no convection and no molecular absorption ($T_{\rm eff}>$9~000~K).
Present version assumes also LTE. NLTE calculations are not yet included into model computations.
There are several simplifying assumptions reducing accuracy of modelling.
Multiple light scattering is treated using simple $\Lambda$-iterations.
A simplified treatment of Stark broadening of H and He lines has been used.
Problems are also an instability of algorithms near Eddington limit and
incompleteness of atomic data.

Typical running times on a PC with CPU 3.2 GHz and 2 GB RAM are several hours
for model atmosphere computation with spectral resolution 300~000 and 64 layers of\break
atmosphere.
One time step in evolutionary computations of separation of mercury isotopes in atmospheres of
CP stars with resolution 5 000 000 takes approximately 15 min.

\section{Special tasks}
SMART enables to compute evolution of chemical composition in atmospheres of CP stars
due to diffusive separation of elements and isotopes driven by radiative acceleration,
light-induced drift and gravity \cite{nachrichten,mons}.
Extensive high-precision line list and collision cross-sections are necessary.
Currently the calculations have been made for mercury isotopes, similar calculations for calcium are in preparation.

Relaxational formation of NLTE in spectral lines can be calculated by rapidly converging iterations.
Equilibrium quantum state populations of ion states are found from the  equations of unbalanced statistical equilibrium
 treated as an relaxational initial value problem from LTE to NLTE populations.

Detailed spectral limb darkening has been computed for some model stellar atmospheres
and used thereafter for finding spectra of rotating stars and non-irradiated eclipsing binaries.
Codes also accounting for stellar surface distortion and gravitational darkening are currently prepared.

To enlighten the problem of stellar wind triggering in stellar atmospheres, the radiative acceleration of
moving clumps with Doppler shifted spectral lines has been studied and found to give hopeful results.

Pan-spectral method for determining element abundances \cite{aveiro} aimed for the
automatic processing of high-quality stellar spectra has been elaborated.
The method is based on weighted cumulative line-widths $Q_\lambda$ defined as
\begin{displaymath}
      Q_\lambda=\int_{\lambda_0}^\lambda{\left|dR_\lambda \over dZ\right|}(1-R_\lambda)d\lambda~,
\label{sap:Q}
\end{displaymath}
where $R_\lambda$ is the residual flux (intensity) and $Z=\log(N_{elem}/N_{tot})$ is the abundance of studied element or isotope.
The derivative of residual flux $R_{\lambda}$ with respect to abundance $Z$ automatically
excludes spectral regions insensitive to changes of the abundance of studied element and
gives a large contribution in the most sensitive regions, i.~e. in the centres of non-saturated
lines and in the steep wings of strong lines of the element.
Best fit of quantities $Q_{\lambda}$ found from synthetic and observed spectra
gives final abundance taking duly into account all lines of studied element including blended ones.
Abundances can be found simultaneously for many elements.
The method can also be used to find corrections of effective temperature and gravity.


\begin{acknowledgments}
We are grateful to Estonian Science Foundation for financial support by grant ETF~6105.
\end{acknowledgments}

\end{document}